\documentstyle[pre,aps,multicol,epsf]{revtex}
\def\B.#1{{\bbox{#1}}} \def\C.#1{{\cal{#1}}}
\begin{document}
\title{Quasi-solitons and asymptotic multiscaling in shell models of
turbulence } \author{Victor S. L'vov} \address{Dept. of Chemical
Physics, The Weizmann Institute of Science, Rehovot 76100, Israel}
\maketitle
%%%%%%%%%%%%%%%%%%%%%%

\begin{abstract} 
A variation principle is suggested to find self-similar solitary
solutions (reffered to as {\em solitons}) of shell model of
turbulence.  For the Sabra shell model the shape of the solitons is
approximated by rational trial functions with relative accuracy of
$O(10^{-3})$. It is found how the soliton shape, propagation time
$t_n$ (from a shell \# $n$ to shells with $n\to \infty$) and the
dynamical exponent $z_0$ (which governs the time rescaling of the
solitons in different shells) depend on parameters of the model. For a
finite interval of $z$ the author discovered {\em quasi-solitons}
which approximate with high accuracy corresponding self-similar
equations for an interval of times from $-\infty$ to some time in the
vicinity of the peak maximum or even after it.  The conjecture is that
the trajectories in the vicinity of the quasi-solitons (with
continuous spectra of $z$) provide an essential contribution to the
multiscaling statistics of high-order correlation functions, referred
to in the paper as an {\em asymptotic multiscaling}. This contribution
may be even more important than that of the trajectories in the
vicinity of the exact soliton with a fixed value $z_0$. Moreover there
are no solitons in some region of the parameters where quasi-solitons
provide dominant contribution to the asymptotic multiscaling.
\end{abstract}

\begin{multicols}{2}
%%%%%%%%%%%%%%%%%%%%%%%%%%%%% 
\section{Introduction}
\label{s:intro}
 
The qualitative understanding of many important statistical features
of developed hydrodynamic turbulence (including anomalous scaling) may
be formulated within the Kolmogorov-Richardson cascade picture of the
energy transfer from large to small scales. For a dynamical modeling
of the energy cascade one may use the so-called shell models of
turbulence \cite{Gledzer,GOY,Jensen91PRA,Piss93PFA,Benzi93PHD,sabra}
which are simplified versions of the Navier-Stokes equations.  In
shell models the turbulent velocity field $\B.u (\B.k ,t )$ with wave
numbers $\B.k $ within a spherical shell $k_n<k< k_{n+1}$ is usually
presented by one complex function, a ``shell velocity'' $u_n(t)$. To
preserve scale invariance the shell wave-numbers $k_n$ are chosen as a
geometric progression
 \begin{equation}\label{kn} k_n=k_0 \lambda^n\,, 
\end{equation} 
where $\lambda$ is the ``shell spacing" and $1\le n \le N$.  The
equation of motion reads $d\, u_n(t)/dt=Q_n$, where $Q_n$ is a
quadratic form of $u_m(t)$ which describes interaction of neighboring
shells.  Clearly, shell models can be effectively studied by numerical
simulations in which the values of the scaling exponents can be
determined very precisely. Moreover, unlike the Navier-Stokes
equations, the shell models have tunable parameters (like $\lambda$)
affecting dynamical features of the of the energy transfer. This
allows one to emphasize one after another different aspects of the
cascade physics and to study them almost separately.

The statistics of $u_n$ may be described by the moments of the
velocity $S_p(k_n)$ which are powers of $k_n$
\begin{equation}
\label{scaling}
S_p(k_n)\equiv \langle |u_n|^p\rangle \propto k_p^{-\zeta_p} \propto
\lambda^{-n \zeta_p}\,,
\end{equation}
in the ``inertial range" of scales, $n_L<n<n_d$.  Here $n_L$ is the
largest shell index affected by the energy pumping and $n_d$ is the
smallest shell index affected by the energy dissipation.

In the paper we employ our own shell model called the Sabra
model\cite{sabra}. Like in the Navier-Stokes turbulence the scaling
exponents $ \zeta_p$ in the Sabra model exhibit non linear dependence
on $p$. Similar anomalies were found in the Gledzer -- Okhitani --
Yamada (GOY) model \cite{Gledzer,GOY}. However  the Sabra model has much
simpler correlation properties, and much better scaling behavior in
the inertial range.  The equations of motion for the Sabra model read:
\begin{eqnarray} \label{sabra}
\frac{d u_n}{dt}&=&i\big( ak_{n+1}  u_{n+2}u_{n+1}^*
 + bk_n u_{n+1}u_{n-1}^*  \\ \nonumber
&& -ck_{n-1} u_{n-1}u_{n-2}\big)  -\nu k_n^2  u_n +f_n\,, \\ \label{abc}
&& a+b+c=0 \ .
\end{eqnarray}
Here
 the star stands for complex conjugation, $f_n$ is a forcing term
which is restricted to the first shells and $\nu$ is the ``viscosity".
Equation~(\ref{abc})  guarantees the conservation of the ``energy" $E$
and ``helicity'' $H$
\begin{equation}\label{energy}
E=\sum_n |u_n|^2\,, \ H=\sum_n (a/c)^n |u_n|^2\,,
\end{equation} 
in the inviscid ($\nu=0$) limit. \\~~\\~~\\

In this paper we will consider self-similar solutions of the Sabra
shell models~(\ref{sabra}) in a form of solitary peaks -- {\em
solitons}.  The important role of intense self-similar solitons in the
statistics of high order structure function was discussed in
Refs~\cite{78Sig,85NN,88Nak}.  The two-fluid picture of turbulent
statistics in shell models and corresponding ``semi-qualitative''
theory in the spirit of Lipatovs' semi-classical approach\cite{76Lip}
was suggested in Refs.~\cite{97DG,00DDG}: self-similar solitons form
in and propagate into a random background of small intensity generated
by a forcing which has Gaussian statistics and $\delta$-correlated in
time. Accounting in the Gaussian approximation for small fluctuations
around self-similar solitons the authors of~\cite{97DG,00DDG} reached
miltiscaling statistics with a narrow spectrum of $z$. In the present
paper the multiscaling statistics of high order correlation functions
will be referred to as {\em asymptotic multiscaling}.

Our preliminary direct numerical simulations of the Sabra shell
model\cite{01LPP,01Pom} shows  the asymptotic multiscaling is a
consequence of much reacher dynamics of shell models.  For example
in the $b$-interval $[-1<b<-0.7]$ and at $\lambda=2$ we indeed observe
extremely intense  self-similar peaks on a background of small
fluctuations. Each particular peak has well defined time-rescaling
exponent $z_0$, however from a peak to peak the value $z_0$
essentially varies\cite{01Pom}.  In the region $[-0.3<b<0]$ the level
of intermittency is much smaller,intense solitary events vanish,
however turbulent statistics remains anomalous\cite{01Pom,00Bif}.
Only at intermediate value of $b\ \approx -0.5$ we found\cite{01LPP}
intense  self-similar peaks with a narrow spectrum of dynamical
exponent $z_0= 0.75 \pm 0.02$.

These observations may serve as a starting point in
developing a realistic statistical theory of asymptotic multiscaling
which will take into account a wide variety of relevant dynamical
trajectories of the system not only in the vicinity of the well
defined solitons. The present paper is a first step in this direction
and is organized as follows.

The analytic formulation of the problem is presented in
Sect.~\ref{s:theory}.  For the general reader I describe a
self-similar form of solitary soliton ``propagating'' through the
shells (\ref{ss:soliton}). I derive the ``basic self-similar
equations'' for the solitons (\ref{ss:equations}), consider the
relevant boundary conditions (\ref{ss:bound}) and analyze the
asymptotic form of soliton tails for infinite times (\ref{ss:tails}).

Section~\ref{s:var} is devoted to a variation procedure for the
problem. I suggest in sect.~\ref{ss:fun} a simple positive definite 
functional $\C.F(z) \ge 0$ such that the exact solution corresponds to
$\C.F(z_0) =0$. The analytic form of trial functions is discussed in
sect.~\ref{ss:whole}. In section~\ref{ss:test} we will see in details
how the variation procedure works in the case of the ``canonical'' set
of parameters $\lambda=2$, $b=-0.5$. The characteristic value of
$\C.F(z)$, $\C.F_0$, is of the order of unity.  The minimization of
$\C.F(z)$ (with respect of the propagation time and soliton width,
with proper choice of trial function without fit parameters and at
experimentally found value $z_0=0.75$\cite{01LPP}) gives $\C.F_{\rm
min}(0.75)\approx 0.08$. Step-by-step improvement of the approximation
is reached by a consecutive addition of fit parameters which affect a
shape of the soliton.  With 10 shape parameters, the value of
$\C.F_{\rm min}(z)$ may be as small as $10^{-3}$.  The resulting 
 ``best'' values of the shape parameters give approximate
solutions of the basic equation (normalized to unity in their maximum)
with local accuracy of the order of $10^{-3}$.

Section~\ref{s:solitons} presents results of the minimization on trial
rational functions (ratios of two polynomials) with 10 shape
parameters, and their discussion. Firstly, in sect.~\ref{ss:z-valley}
we compare and discuss the $z$-dependence of $\C.F_{\rm min}(z)$ for
$b=-0.8\,,\,-0.5$ and $-0.3$ (at $\lambda=2$). An important
observation (sect.~\ref{ss:quasi}): there are interval of $z$ for
which basic equations do not have self-similar solitary solutions for
all times (solitons) but may be solved approximately for interval of
times from $-\infty$ up to some moment in the vicinity of soliton
maximum or even after it.  Configurations of the velocity field in the
vicinity of these solutions may be called {\em quasi-solitons}. My
conjecture is that the quasi-solitons with continuous spectra of $z$
may provide even more important contribution to the asymptotic
multiscaling than the contributions from the trajectories in the
vicinity of an exact soliton with fixed scaling exponent $z_0$. The
concept of quasi-solitons and proposed in this paper the dependence of
their properties on $b$ allows us to reach a qualitative understanding
of the behavior of intense events for various values of $b$, observed
in direct numerical simulation of the Sabra
model\cite{01LPP,01Pom,00Bif}.

 In concluding sect.~\ref{s:concl} I summarize the results of the
 paper and present my understanding of a way ahead toward a realistic
 theory of asymptotic multiscaling for shell models of turbulence
 which also may help in further progress in the description of
 anomalous scaling in the Navier-Stokes turbulence.

%%%%%%%%%%%%%%%%%%%%%%~(\ref{})%%%%%%%%%%%%%
\section{Basic  Self-similar Equations of the Sabra shell  model}
\label{s:theory}

%%%%%%%%%%%%%%%%%%%%%%%%%%%%%%%%%%%%%%%%%%%%%%%%%%%%%%%%%%%%
\subsection{``Physical'' range of parameters}
\label{ss:range}
In the inertial interval of scales the Sabra equation of
motion~(\ref{sabra}) have formally five parameters: $k_0$, $\lambda$,
$a$, $b$ and $c$. They enter in the equation in four combinations:
$\lambda$, $(k_0\,a)$, $(k_0\,b)$ and $(k_0\,c)$, therefore by
rescaling of the parameters $a,\,b$ and $c$ we get $k_0=1$.  Without
loss of generality we may consider $a >0$. A model with negative $a$
turns into model with positive $(k_0\,a)$ by replacing $u_n\to -u_n$.
By rescaling of the time-scale $t\to (a\,t)$ we get a model with
$a=1$. The arameters $a$, $b$ and $c$ are related by
Eq.~(\ref{abc}). Thus we can express $ c=-(a+b)\to -(1+b)$.   With this
choice only two parameters of the Sabra model remain independent, $b$
and $\lambda$.  By construction of the model $\lambda>1$. A typical
choice $\lambda=2$ will be considered in the paper.

Note that for $(a/c)>0$ (which is $c>0$ or $b<-1$ at $a=1$) the model
has two positive definite integrals of motion which are quadratic in
$u_n$: the energy and ``helicity''~(\ref{energy}).  In this case (as
it was discussed in Ref.~\cite{95BK}) one may directly apply the
Kraichnan argument for the enstropy and energy integrals of motion in
2D turbulence and conclude that in shell models fluxes of energy and
``helicity'' will be oppositely directed: ``direct'' cascade (from
small to large shell numbers $n$) will have integral of motion for
which large $n$ shells will dominate. Therefore for $(a/c)>1$ (which
is $-2<b<-1$ at $a=1$) one expects direct cascade of ``helicity'' and
inverse cascade of energy, like in 2D turbulence. This reasoning
predicts direct cascade of energy for $(a/c)<1$ (or $b<-2$ at $a=1$).
In both cases one cannot expect a statistically stationary turbulence
with flux equilibrium because one of the (positive definite)
integrals of motion will accumulate on first shells without a mechanism
of dissipations. Therefore one expects an energy-flux equilibrium with
direct cascade of energy like in 3D turbulence only for negative ratio
$a/c$ (or $b>-1$ at $a=1$) when ``helicity'' integral is not
positive definite and Kraichnan's arguments are not applicable.

More careful analysis shows that only the region $-1<b<0$ (at $a=1$)
may pretend to mimic 3D turbulence. For $b>0$ when $(-c)>a=1$ the
existence of ``helicity'' integral (even not positive definite) leads
to a period-two oscillations of the correlation functions which are
increasing with $n$ and ``unphysical'' from the viewpoint of 3D
turbulence.  One can see this from the exact solution for the
3rd-order correlation function
\begin{equation}\label{eq:S3}
  S_3(k_n)\equiv\mbox{Im}\left<u_{n-1}u_n u_{n+1}^*\right>\,,
  \end{equation}
which in the inertial interval of scales reads\cite{sabra}:
\begin{equation}\label{eq:S31}
S_3(k_n)=\frac{1}{2k_n(a-c)}\Big[-\bar\epsilon + \bar\delta
\Big(\frac{c}{a}\Big)^n\Big]\ .
\end{equation}
Here $\bar\epsilon$ and $\bar\delta$ are fluxes of energy and
``helicity'' respectively. For any small ratio
$\bar\delta/\bar\epsilon$ and $(c/a) < -1$ the correlator $S_3(k_n)$
(and presumably many others) will have period-two oscillations which
increase with $n$  . Therefore in this paper we will consider only the
region
$$
-1<b<0\,,
$$ traditionally keeping $\lambda=2$.

\subsection{Self-similar form of ``propagating'' solitons}
\label{ss:soliton}
Self-similarity in our context means that solitons propagate through
shells {\em without changing their form}.  ``Propagation'' means that
the time $t_n$ at which peak reaches its maximum increases with $n$,
in other words, the larger $n$, the later peak reaches this
shell. Intuitively this picture corresponds to the energy transfer
from shells with small $n$ to large ones.

Self-similar propagation of the solitons may be formally described by
the same function $f(\tau_n)$ of dimensionless time $\tau_n$ which is
counted from the time of the soliton maximum $t_n$ and normalized by
some characteristic time for $n$th shell $T_n$:
\begin{equation}
\label{eq:taun}
    \tau_n=(t-t_n)/T_n\ .
  \end{equation}
The time $T_n$ has to be rescaled with $n$ as follows:
\label{eq:time-rescale}
  \begin{equation}
T_n=T  \lambda^{-zn}\,,\quad T=1/(k_0\,v)\,,
\end{equation}
where $z$ is a {\em dynamical exponent} and the characteristic time
$T$ is organized from the characteristic velocity of the soliton $v$
and the wave vector of the problem $k_0$.  The time delays
$t_n-t_{n-1}$ also have to re-scale like $T_n$:
\begin{equation}
\label{times}
t_{n-1,n}\equiv t_n - t_{n-1}= \tilde \tau T\lambda ^{-zn}\,,
\end{equation}
where the {\em positive} dimensionless time $\tilde \tau $ is of the
order of unity.

Consider next the amplitudes of the solitons. Denote as $u_{n,\rm max}$
the maximum of the velocity in the $n$th shell which also  re-scales
with another exponent $y$:
\begin{equation}
  \label{eq:am-rescale}
  u_{n,\rm max}=v \lambda^{-yn}\ .
\end{equation}
In order to relate the exponents $z$ and $y$ we sketch the basic
equations~(\ref{sabra}) having in mind only dimensions and $\lambda^n$
dependence:
$$
 (du_n/dt)\propto k_0\,\lambda^n  u_n^2\ .
$$ Consequently $ T_n^{-1}\sim k_0 \lambda ^n u_{n,\rm max}$ \ and
therefore
\begin{equation}
  \label{eq:y-z}
  y+z=1\ .
\end{equation} 
Finally we may write a self-similar substitution in the form:
\begin{equation}
u_n(t)= -i v\lambda^{-yn} f\left[(t-t_n)vk_0 
\lambda^{zn} \right]\ .
\label{scform1}
\end{equation}
With this choice of a prefactor $(-i)$, the real and positive function $f$
will give a positive contribution to the energy flux~(\ref{eq:S3}) in
the inertial interval of scales:
\begin{equation}
  \label{eq:flux1}
  \bar\epsilon=2 k_n\, (a-c)\, \mbox{Im}\langle u_{n-1}^* u_n^*
  u_{n+1}\rangle \ .
\end{equation}
Note that all these considerations are not specific for the Sabra
model. They are based on very general features of shell models of
turbulence, namely the quadratic form of nonlinearity with the amplitudes
of interaction proportional to $ \lambda^n$.  In
Refs.~\cite{88Nak,97DG} similar forms of the self-similar substitutions
were taken for the Obukhov -- Novikov  and for the GOY models.

\subsection{Self-similar equation of motion}
\label{ss:equations}
Introduce a dimensionless time for the $n$th shell as follows:
\begin{equation}
 \label{tau-n}
  \tau_n \equiv (t-t_n)vk_0  \lambda^{zn}\ .
\end{equation}
The right-hand side (RHS) of the equation for $d \,f(\tau_n)/d\,\tau_n
$ will involve a function $f$ with arguments $\tau_{n\pm 1}$ and
$\tau_{n\pm 2}$.  All these times may be uniformally expressed in
terms of a new dimensionless time
\begin{equation}
  \label{eq:tau-0}
  \tau_0\equiv \tilde \tau / (\lambda^z-1)
\end{equation}
as:
\begin{equation} \label{times1}
  \tau_{n\pm s}=\lambda^{\pm s z}(\tau_{n}-\tau_0)+\tau_0\,,\quad  
s=1,\, 2\ .
\end{equation}
 The characteristic time $\tau_0$ is related to the time $t_{n,\infty}$
 which is needed for a pulse to propagate from the $n$th shell all the
 way to infinitely high shells:
 \begin{equation}\label{eq:tn0}
t_{n,\infty}\equiv \sum_{m=n}^{\infty}t_{m,m+1}= \tau_0 \lambda
   ^{-nz}\,  T \ .
 \end{equation}
Substituting Eq.~(\ref{scform1}) in~(\ref{sabra}) and using
relationship~(\ref{times1}) one gets the ``basic equation'' of our
problem
\begin{equation}\label{eq:eq}
D(\tau)\equiv \frac{d\, f(\tau)}{d\tau}- C(\tau)\ .
\end{equation}
Here we replaced $\tau_n\to \tau$ and introduced a ``collision'' term
$C(\tau)$ according to:
\begin{eqnarray}\nonumber 
&&C(\tau) \equiv -a\,\lambda
^{3z-2}\,f^*[\lambda^{z}(\tau-\tau_0)+\tau_0]\, f[\lambda^{2
z}(\tau-\tau_0)+\tau_0]\\ \nonumber &&-\,\,c\lambda ^{2-3z\,}f
[\lambda^{-z }(\tau-\tau_0)+\tau_0]\, f[\lambda^{-2
z}(\tau-\tau_0)+\tau_0]\\ &&+(a+c)\,f^*
[\lambda^{-z}(\tau-\tau_0)+\tau_0]\, f
\,[\lambda^{z}(\tau-\tau_0)+\tau_0]\ .\label{st}
\end{eqnarray}
%%%%%%%%%%%%%%%%%%%%%%%%%%%%%%%%%%%%%%%%%%%%%%%%%
\subsection{Boundary conditions of the basics equation~(\ref{eq:eq})}
\label{ss:bound}
The boundary conditions at $\tau=\pm\infty$ for a soliton are obvious:
 \begin{equation} 
   \label{eq:bound-inf}
f(\pm\infty)=0 \ .
 \end{equation}
By construction the soliton reaches a maximum at $\tau=0$. Therefore
\begin{equation} 
   \label{eq:bound1}
\frac{df (\tau)}{d\tau}\Big|_{\tau=0}=0\ .
 \end{equation}
Introduce a characteristic width of a soliton $1/d$ according to
\begin{equation}
  \label{eq:d}
 \frac{d^2\, f(\tau)}{d\,\tau^2}\Big|_{\tau=0}=- d^2\ .
\end{equation}
It is convenient to introduce a new time variable $s$ and a function
$g(s)$ with the unit width:
\begin{eqnarray}
  \label{eq:gs}
&& f(\tau)\equiv g(s)\,,\quad s\equiv \tau\,d \,,\\
 &&   \frac{d^2\, g(s)}{d\,s^2}\Big|_{s=0}=-1\ .
\end{eqnarray} 
With this time variable, the  equation of motion~(\ref{eq:eq}) reads
\begin{equation}\label{eq:eq-s}
\C.D(s,z,s_0,d) = q(s) - \C.C(s,z,s_0)=0\,, 
\end{equation}
where
\begin{eqnarray}\label{eq:phi}
 &&~ \qquad q(s)\equiv d\,\frac{d\, g(s)}{ds}\,, \qquad s_0\equiv
 \tau\,d\,,
\\ \nonumber &&\C.C(s,z,s_0) \equiv (a+c)\,g^*
 \big(\frac{s-s_0}{\lambda^{z}}+s_0\big)\, g
 \,[\lambda^{z}(s-s_0)+s_0]\\ \nonumber &&-a\,\lambda ^{3z-2}\,
 g^*[\lambda^{z}(s-s_0)+s_0]\, g[\lambda^{2 z}(s-s_0)+s_0]\\
 &&-\,\,c\lambda ^{2-3z\,}g \big(\frac{s-s_0}{\lambda^{z}}+s_0\big)\,
 g\big(\frac{s-s_0}{\lambda^{2z}}+s_0\big)\,, \label{eq:coll}
\end{eqnarray}
Function $g(s)$ should vanish at infinite times:
 \begin{eqnarray}
   \label{eq:infty-}
 \lim_{s\to -\infty} g(s)&=& 0\,,\\ \label{eq:infty+} 
\lim_{s\to   \infty} g(s)&=& 0\ .
 \end{eqnarray}
The boundary conditions at $s=0$ read:
\begin{equation}
\label{eq:max}
g(0)=1\,, \quad \frac{d\, g(s)}{d\,s}\Big|_{s=0}=0\,,\quad \frac{d^2\,
g(s)}{d\,s^2}\Big|_{s=0}=-1\ .
 \end{equation}
Note that the problem to find a form of the self-similar solitons may
be divided into two independent problems: for times smaller and larger
than $s_0$.  Indeed, Eq.~(\ref{eq:eq-s}) for times $s < s_0$ does not
contain functions for times $s > s_0$ and {\em vice versa}.  At the
boundary between these regions Eq.~(\ref{eq:eq-s}) reduces to
\begin{eqnarray}\label{eq:f1}
\frac{d }{d\,s_0}\Big[\frac{1}{g(s_0)}\Big] &=& 
d\, p_0\,, \\ \nonumber
p_0&=& a (\lambda^{3z-2}-1) - c(1-\lambda^{2-3z}) \ .
\end{eqnarray}
In our discussion $a>0$, $c<0$, $\lambda>1$ and $z>2/3$. Therefore
$p_0>0$ and $ d \, g(s_0)/ d\,s_0<0$. It means that the time $s_0$ is
larger than the time $s=0$ at which $g(s)$ has a maximum, {\em i.e.}
$s_0>0$.  As we noted [and see Eq.~(\ref{eq:tn0})], $s_0$ is the time
which is needed for a pulse to propagate from the $n$th shell to
infinity high shell. Therefore the relation $s_0>0$ agrees with our
understanding of the direct cascade.

So, we will divide the time interval $[-\infty <s<\infty]$ into two
subintervals: $[-\infty <s\le s_0]$ and $[s_0 <s<\infty]$.  For
$[-\infty <s\le s_0]$ the maxima of the solitons in all shells are in
the inertial interval of scales and very high shells are not yet
activated.  We will refer to this interval as an {\em inertial
interval of times}. In the second time interval, for $s>s_0$
high-shell solitons already reached the dissipative interval of
scales.  We will refer to this interval as a {\em dissipative interval
of times}. Generally speaking, in this interval one has to account for
the viscous term in the equation of motion.  In this paper we will
restrict ourselves to the inertial interval of times.

It was shown in Ref.~\cite{88Nak} that equations similar to
(\ref{eq:eq}) with similar boundary conditions can be considered as 
nonlinear eigenvalue problems. They have trivial solutions $f(s)=0$,
but they may have nonzero solutions for particular the values of
$z=z_0$ and $s_0$.  Below we will find nontrivial solutions of our
Eq.~(\ref{eq:eq-s}) that satisfies conditions (\ref{eq:infty-}),
(\ref{eq:max}) and for which $z_0$ lies in the physical region
$\case{2}{3}<z<1$.

 \subsection{Qualitative analysis of self-similar solutions 
at $s \to \pm \infty$}
\label{ss:tails}
 In this subsection we will analyze time dependent solutions
 of~(\ref{sabra}) which are more general than a Kolmogorov-41 (K41)
 and have a form of solitary pulses -- solitons.  We will show that
 the solitons have long (power-like) tails $g(s) \propto s^{x_\pm}$ at
 $s \to \pm \infty$. One way to find the asymptotic solution of
 Eq. ~(\ref{eq:eq-s}) is to balance the exponents in its left-hand
 side (LHS) and RHS.  For $g(s)\propto s^{-x}$ this gives immediately
 $x=1$. Thus: $ g(s)= D_1/s$.  Equating prefactors in the LHS and RHS
 of the Eq.~(\ref{eq:eq}) we get:
\begin{equation}
\label{1nag}
D_1=\frac{-  \lambda^2\, d }{(a-c\lambda^2)(\lambda^2-1)}\ .
\end{equation}
The coefficient $D_1$ appeared to be real and, in the actual range of
the parameters~($\lambda>1\,\ a>0\,,\ c<0$), negative.  For the
positive function $g(s)$ this asymptotic form describes the front part
of the pulse (at negative $s$).  So:
\begin{equation}\label{asympt2}
g(s)= \frac{-\lambda^2\,d }{s (a-c\lambda^2) (\lambda^2-1)} \,, \quad
s \to -\infty\ .
\end{equation}
Another approach is to assume that in the equation $[g(s)\propto
s^{-x}]$ the exponent $x<1$.  Then, the LHS will behave as $1/s^{1+x}$
while the RHS will be proportional to $1/s^{2x}$. In the limit $s\to
\infty$ and at $x <1$ one may neglect the LHS of the
Eq.~(\ref{eq:eq-s}).  Then the exponent $x$ cannot be found by
power-counting.  Instead one requires that the prefactor in the RHS
[{\sl i.e.} in the $\C.C(s)$ term] must vanish. This gives the
following equation for $x$:
\begin{equation}
\label{order1}
0=a\lambda^{3z(1-x)-2}+c\lambda^{2-3z(1-x )}-(a+c) \ .
\end{equation}
Denoting 
\begin{equation}\label{defL}
\Lambda\equiv\lambda^{3z(1-x)-2}\,,
\end{equation}
we have, instead of Eq.~(\ref{order1}), the square equation for
$\Lambda$:
$$
a\Lambda^2-(a+c)\Lambda+c=0\,,
$$
with the roots
$$ 
\Lambda_1=1 \,, \ \ \Lambda_2 = c/a \ .
$$ 
In the chosen region of parameters ($c$ and $a$ have different signs)
$\Lambda_2<0$, which contradicts the assumption of real $x$.  Therefore
the only root $\Lambda_1=1$ is relevant.  According to
definition~(\ref{defL}) this gives
\begin{equation}
\label{rel}
x_1 =\gamma\equiv 1-\frac{2}{3z}\ .
\end{equation}
 Here we used notation $\gamma$ for the exponent of the long positive
 tail of the pulse introduced in\cite{88Nak} for the Novikopv-Obukhov
 shell model.  As we see, the relation~({\ref{rel}}) is model
 independent.  Actually this equation is a consequence of the
 conservation of energy (reflected in the constraint $a+b+c=0$) and
 the fact that the shell models account only for interaction of three
 consequent shells.  Clearly, the relevant region of $\gamma$ is
 $0<\gamma<1$. This corresponds to
\begin{equation}\label{regionz}
{  2 \over 3}<z< 1\ .
\end{equation}  
We conclude that for $s\to +\infty$
\begin{equation}
  \label{eq:infty}
  g(s)=D_\gamma/s^\gamma \,,
\end{equation}
with a free factor $D_\gamma$ which has to be determined by matching
the asymptotics~(\ref{eq:infty}) with a solution in the region $s\sim
s_0$.  It is known~\cite{85NN} that the self-similar core of the
function $g(s)$ gives rise to a linear $n$-dependence of the scaling
exponent $\zeta_n$ of the $n$th order structure function at large $n$:
\begin{equation} \label{zn}
\zeta_n=z+(1-z)n\,, \quad \mbox{at}\ n>n_{\rm cr}=1/\gamma\ .
\end{equation}

For $z=\case{2}{3}$ Eq.~(\ref{eq:infty}) gives the K41 slope
$\case{1}{3}$ which is the largest possible one.  Value $z=1$
corresponds to the largest possible intermittency (zero
slope). Therefore we may consider Eq.~(\ref{regionz}) as the ``physical
region'' of the dynamical exponent $z$.
%%%%%%%%%%%%%%%%%%%%%%%%%%%%%%%%%%%%%%%%%%%%%%%%%%%%%%%%%%%%%
\section{Variation procedure for solitons and quasi-solitons}
\label{s:var}
\subsection{Suggested functionals}
\label{ss:fun}
Consider the following functional acting on the function $g(s)$
\begin{equation}
  \label{eq:fun}
 \C.F\{g(s)|z,s_0,d\}\equiv \sqrt{\int_{-\infty}^{s_0} d\,s
 [\C.D(s,z,s_0,d)]^2}\,,
\end{equation}
which also depend on the parameters of the problem $\lambda$, $b$,
$z$, $s_0$ and $d$. By construction, the functional
$\C.F\{g(s)|z,s_0,d\}$ is non-negative and equal zero if $g(s)$ is a
solution of the problem~(\ref{eq:eq-s}):  $\C.D=0$.   Clearly, the
functional~(\ref{eq:fun}) is not unique. For example, we may use more
general, ``weighted'' functional,
\begin{equation}
  \label{eq:tfun}
 \tilde \C.F\{g(s)|z,s_0,d\}\equiv \sqrt{\int_{-\infty}^{s_0} d\,s
W(s) [\C.D(s,z,s_0,d)]^2}\,,
\end{equation}
with some positive ``weight'' function $W(s)>0$. This functional is
also positive definite and also equal zero if $g(s)$ is a solution of
the problem~(\ref{eq:eq-s}).

One can easily find many other functionals giving approximate
solutions of the problem.  The functional~$\C.F\{g(s)|z,s_0,d\}$ has
an advantage of simplicity. Its minimization (with a proper choice of
the trial function) leads to a solution of the problem with hight
accuracy. For example, with trial functions discussed in
Section~\ref{ss:whole} the relative accuracy (with respect to the
value of a soliton maximum) is $O(10^{-3})$. Therefore we will
restrict the present discussion to the simple functional
$\C.F\{g(s)|z,s_0,d\}$.

%%%%%%%%%%%%%%%%%%%%%%%%%%%%
\subsection{Suggested  form of trial functions}
\label{ss:whole}
 For simplicity, in this paper we will seek only real (without
 nontrivial phases) solitons. Complex solitons will be discussed
 (within the same scheme) elsewhere.

My suggestion is to use different trial functions for negative and
positive times, both satisfying boundary conditions~(\ref{eq:max}) at
$s=0$.  Denote as $g_{m+}(s)$ the trial functions for positive times
with $m$ shape parameters.  Let the function $g_{m+}(s)$ satisfy
condition~(\ref{eq:f1}) at $s=s_0$. This condition is a constraint on
the time derivative of $1/g_{m+}(s)$.  Having also in mind that
$g_{m+}(s)$ is defined on the interval $[0<s<s_0]$ with $s_0\sim 1$ it
is convenient to choose $1/g_{m+}(s)$ as a polynomial in $s$. The
first (free) term of expansion is 1 because $g_{m+}(0)=1$. The second
term ($\propto s$) vanishes due to $d g(s)/ds =0$ at $s=0$. The next
term must be $s^2/2$ due to the restriction on the second
derivative~(\ref{eq:max}) at $s=0$. In order to satisfy
condition~(\ref{eq:f1}) at $s=s_0$ we have to account at least for a
cubic term. Therefore the simplest function of this type, $g_{0+}(s)$
without fit parameters takes the form
\begin{equation}
  \label{eq:fit+0}
  g_{0+} (s)\equiv
\Big[1+\frac{s^2}{2}   +\frac{s^3}{3s_0^2}\Big(\frac{p_0}{d}-s_0
\Big)   \Big]^{-1}\ .
\end{equation}
If needed we can add fit parameters $p_1$, $p_2$, {\em etc.}
accounting for terms $\propto s^4$, $\propto s^5$, {\em etc.}
Accordingly, a trial function with $m$ shape parameters takes the
form:
\begin{eqnarray}
  \label{eq:try} g_{m+}
(s)&=& \Big[1+\frac{s^2}{2}+\frac{s^3}{3s_0^2}\Big(\frac{p_0}{d}-s_0
-\sum_{j=1}^{m}p_j\Big) \nonumber \\ && +\sum_{j=1}^{m} \frac{p_j \,
s^{j+3}}{(j+3)\, s_0^{j+2}}\Big]^{-1} \ .  \nonumber
\end{eqnarray} 
In most cases it would be enough to use $g_{5+}(s) $ with 5 shape
parameters.

Denote as $g_{m-}(s)$ the trial functions for negative times having $m$
shape parameters. These functions have to approximate the basic
equation on the infinite time interval from $-\infty$ to $0$.
Therefore their analytic form is a much more delicate issue. Analyzing
the form of the peaks observed in direct numerical
simulations\cite{01Pom}.  I have found a function $\tilde g_{1-}(s)$
with one fit parameter $q$,
\begin{equation}
\label{eq:tfit-1}
\tilde g_{1-} (s)\equiv \frac{1-s \sqrt{q/(q-1)}} {\big[1-s /
\sqrt{q(q-1)}\, \big]^q}\,,
\end{equation}
which allows us to reach accuracy of solution of Eq.~(\ref{eq:eq-s})
of  $O(10^{-2})$.  This accuracy would be enough for
future comparison of the ``theoretical'' shape of a pulse with that
found in direct numerical simulation of the Sabra model.  However this
function does not have the correct asymptotic behavior~(\ref{asympt2}) for
$s\to -\infty$ and is inconvenient for  successive improvements of the
approximation.

Instead of $\tilde g_{1-}(s)$ for negative time I use in the regular
minimization procedure a ratio of two polynomials of $n$th and $n+1$
orders. This ratio has $2n+2$ free parameters. After accounting for
three conditions~(\ref{eq:max}), just $(2n-1)$ parameters remain free.
Simple function of this type with $n=1$, which also agrees with
asymptotic~(\ref{asympt2}), has no free parameters:
\begin{equation}
  \label{eq:tfit-0}
\tilde g_{0-}(s)=\frac{2+D_1s}{2+D_1s+s^2}\ .
\end{equation}
We will discuss even more simple rational function
\begin{equation}
  \label{eq:fit-0}
g_{0-}(s)=[1+\case{1}{2}s^2]^{-1}\ .
\end{equation}
Actually a better approximation may be achieved by a function similar
to~(\ref{eq:tfit-0}) with one free parameter $q_1$:
\begin{equation}
  \label{eq:fit-1}
g_{1-}(s)=\frac{1+q_1s}{1+q_1s+\case{1}{2} s^2}\ .
\end{equation}
In actual calculations it would be sufficient for our purposes to
account for 5 shape parameters in the function:
 \begin{equation}
  \label{eq:fit-5}
g_{5-}(s)=\frac{1+ q_1 s+ q_2s^2 + q_4 s^3}{1+ q_1 s+
(q_2+\case{1}{2}) s^2+ q_3 s^3+ q_5 s^4 }\ .
\end{equation}
The function $g_{5-}(s)$ reduces to $g_{3-}(s)$ by choosing $q_4=
q_5=0$ and to $g_{1-}(s)$ at $q_2=q_3=q_4=q_5=0$.
%%%%%%%%%%%%%%%%%%%%%%%%%%%%%%%%%%%%%%%%%%%%%% %%%%%%%%%%%%%%%
\subsection{Test of the   variation procedure}
\label{ss:test}
In this subsection we will see in detail how the variation procedure
suggested above works for the ``canonical'' set of parameters
$\lambda=2$, $b=-0.5$.  Denote as $\C.F_{{\rm min},m}(z)$ the result
of a minimization of the functional $\C.F\{g(s)|z,s_0,d\}$ acting on
the functions $g_{m\pm}(s)$ at given $z$. The parameters in the
minimizations are: the propagation time $s_0$, width parameter $d$,
$m$ shape parameters for negative times ($q_1,\cdots q_m$) and $m$
shape parameters for positive times ($p_1,\cdots p_m$):
\begin{equation}    \label{eq:Fmin}
\C.F_{{\rm min},m}(z) \equiv  \min_{q_j,p_j|s_0,d}
\C.F\{g(s)|z,s_0,d\}\ .
  \end{equation}
Table~1 displays the found values of $\C.F_{{\rm min},m}(z)$ for
$z=0.75,\,0.85$, $0.95$ and $m=0$, 1, 3, 5. 

Consider first the case $z=0.75$, which corresponds to the dynamical
exponent ofintense self-similar peaks observed in our direct numerical
simulation of the Sabra shell model~\cite{01LPP}.  At $m=5$ we have
reached $\C.F_{\rm min,5}(0.75)\approx 2.3\cdot 10^{-4}$.  This value
is more than 200 times smaller than characteristic value of the
functional before minimization, $\C.F_0\approx 0.5$. This allows us to
hope that the minimization using the functions $g_{5\pm}(s)$ with
total number $2\, m=10$ of the shape parameters will be sufficient for
most applications.

Note that for some purposes we may use less then 10 shape
parameters. For example ``optimal'' values of $s_0$ and $d$ shown in
the Table~1 begin to converge for $m\ge 3 $. Therefore for a
reasonable good estimate of $s_0$ and $d$ we may use the trial functions
$g_{3\pm}(s)$ having 6 shape parameters.  Moreover, the same level of
accuracy as with $m=3$ may be achieved just with two shape parameters
($q,\,p_1$) if we replace $g_{3-}(s)$ by $\tilde g_{1-}(s)$ and,
(which is less important) $g_{3+}(s)\to g_{1+}(s)$. Minimization with
$\tilde g_{1-}(s)$ and $g_{1+}(s)$ gives:
\begin{eqnarray}
  \label{eq:50res2}
\min_{s_0,d,q,p_1}\C.F\{g(s)|z,s_0,d\}\approx 0.0197\,,\ \mbox{at}\ z
&=&0.75\,, \\ \nonumber s_0\approx 0.54\,,\ d\approx 1.57\,,\ q\approx
8.59\,,\ p_1&\approx & -0.44 \ .
\end{eqnarray}

\begin{figure}
\epsfxsize=8.3cm \epsfbox{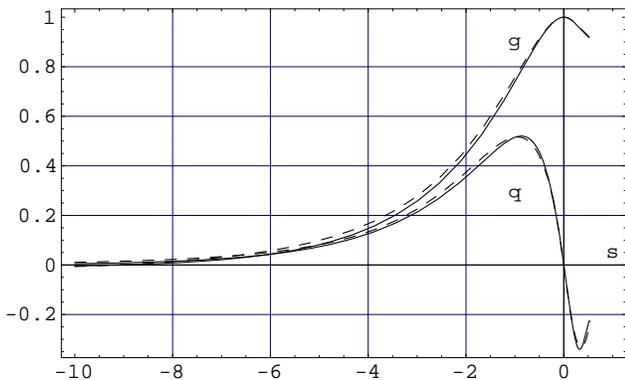}
\caption{Trial functions $g(s)$ and  
$q(s)\equiv d[dg(s)/ds]$ for $\lambda =2$, $b=-0.5$ and
$z=0.75$. Solid lines correspond to the full minimization ($m=5$),
dashed -- to $m=3$.\ The same dashed lines show results of the
minimization~(\ref{eq:50res2}) with only two shape parameters.}
 \label{f:shape1}
\end{figure}

Another example in which we may use less then 10 shape parameters for
a reasonable good description is the shape of solitons, $g(s)$.
Fig.~\ref{f:shape1} displays functions $g(s)$ and $q(s)$,
Eq.~(\ref{eq:phi}) for $m=3$ (dashed lines) and for $m=5$ (solid
lines). We see that for comparison with experiments we can use simpler
form with $m=3$. In Fig.~\ref{f:shape1} are also plotted by dashed
lines functions $g(s)$ and $q(s)$ resulting from
minimization~(\ref{eq:50res2}) with 2 shape parameters. They are
indistinguishable from the corresponding functions having 6 shape
parameters.

%%%%%%%%%%%%%%%%%%%%%%%%%%%%%%%%%%%%%%%%%%%%%%%%%%%%%%%%

Table~1.~Optimal values of $(s_0,\,d)$ for $z=0.75$ and ~values of
$\C.F_{{\rm min},m}$ for $z=0.75,\,0.85$ and $0.95$.  Number of the
shape parameters $m=0,\,1,\,3$ and 5.

\begin{center}
\begin{tabular}{||c|c||c|c|c|c|| } 
\hline \hline & & $~m=0~$& $~m=1~$& $~m=3~$ & $~m=5~$ \\ \hline \hline
$z= 0.75 $~ & $s_0$ & 0.36 & 0.36 & 0.55 & 0.517 \\ \hline $z=0.75 $~
& $~ d$ & 1.19 & 1.18 & 1.59 & 1.49 \\ \hline\hline $z=0.75 $~ &
$\C.F_{{\rm min},m}$ & ~0.0811 & ~0.0675 & 0.0176 & ~~0.00234~ \\
\hline $z=0.85 $~ & $\C.F_{{\rm min},m}$ & ~0.2047 & ~0.1014 & 0.0284
& 0.0092\\ \hline $z=0.95 $~ & $\C.F_{{\rm min},m}$ & ~0.3471 &
~0.1377 & 0.0780 & 0.0650 \\ \hline\hline
\end{tabular}
\end{center}
%%%%%%%%%%%%%%%%%%%%%%%%%%%%%
%%%%%%%%%%%%%%%%%%%%%%%%%%%%%

%%%%%%%%%%%%%%%%%%%%%%%%%%%%%

\begin{figure}
\epsfxsize=8.3cm \epsfbox{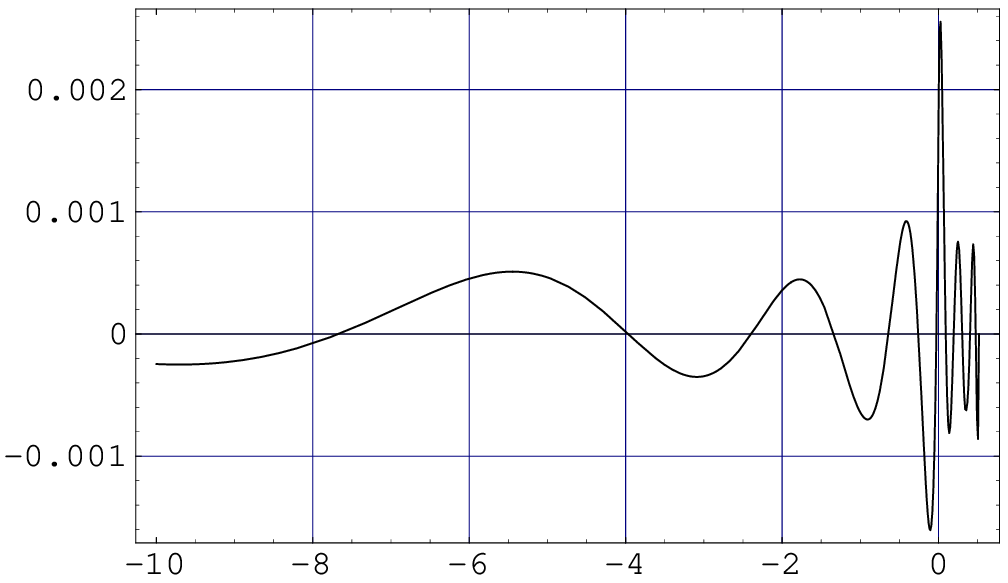}\vskip 0.2cm
\epsfxsize=8.3cm \epsfbox{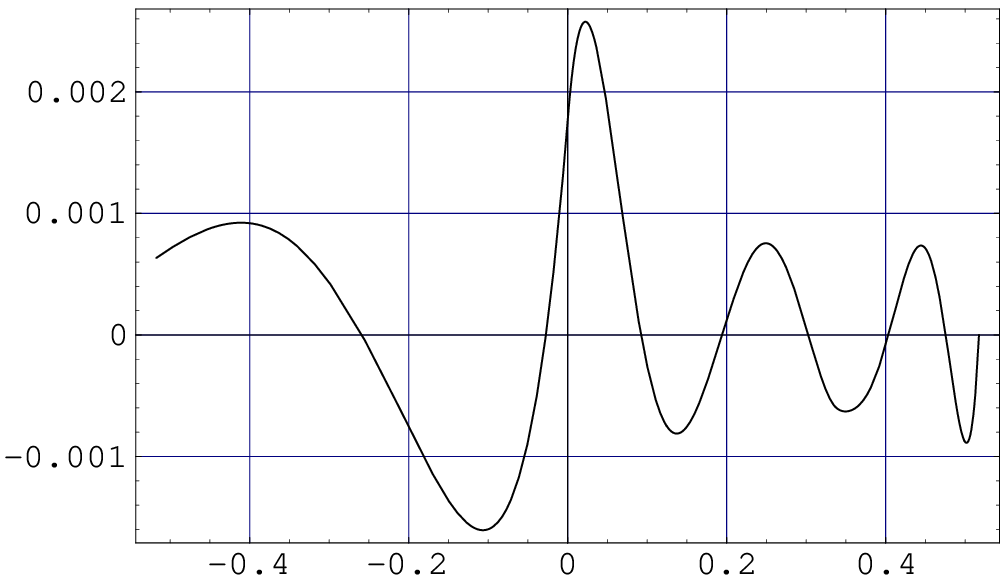}
\caption{Local deviation $\C.D_{{\rm
min},5}(s,0.75)$, Eq.~(\ref{eq:eq-s}) {\em vs} $s$ for $\lambda =2$,
$b=-0.5$. Upper panel: region $-10<s<s_0$. Lower panel: blow up of the
region $-s_0<s<s_0$.}
 \label{f:deviation1}
\end{figure}
%%%%%%%%%%%%%%%%%%%%%%%%%%%%% 
Note that the minimal values of the functional $\C.F_{{\rm min},m}(z)$
characterize the ``global'' accuracy of the approximation in the whole
interval of $s$, [$\infty\,,\ s_0$]. It would be elucidative to
discuss a ``local deviation'' which may be described by the function
$\C.D(s,z,s_0,d)\equiv q(s)-\C.C(s,z,s_0)$.

According to Eq.~(\ref{eq:eq-s}), the local deviation
$\C.D(s,z,s_0,d)$ has to vanish for all $s$. Denote as $ \C.D_{{\rm
min},m}(s,z)$ the deviation $ \C.D(s,z,s_0,d)$ at optimal values of
$s_0,d$ and $2m$ shape parameters. The deviation $\C.D_{{\rm
min},5}(s,0.75)$ is shown in Fig.~\ref{f:deviation1}.  Excluding the
small region $|s|<0.1$ the deviation $\C.D_{{\rm min},5}(s,0.75) <
10^{-3}$, {\em i.e.} in more than 500 times smaller than the maximal
value of $ q(s)$ shown in Fig.~\ref{f:shape1}. This serves for us as a
strong support of the conjecture that by adding more and more fit
parameters one can reach smaller and smaller values of $\C.F_{{\rm
min}}$, {\em i.e.}  $\lim_{m\to \infty}\C.F_{{\rm min},m}=0$ and for
$z=0.75$ one can find a true solution of the problem.  The currently
available personal computers allow to find solutions with $\C.F_{{\rm
min},10}\sim 10^{-3}$ and $\C.D< 2 \cdot 10^{-3}$ during one-two hours
of calculations using the standard package of Wolfram' {\sl Mathematica}.

Consider now different a value of $z=0.95$.  As follows from the
Table~1, there are no jumps in improving the approximation for $m\ge
2$. For $z=0.95$ the ratio $\C.F_{{\rm min,1}}(z)/\C.F_{{\rm
min,5}}(z) \simeq 2$ while for $z=0.75$ the same ratio $\C.F_{{\rm
min,1}}(z)/\C.F_{{\rm min,5}}(z)\simeq 28$. It is very reasonable to
expect that $\C.F_{{\rm min},m}(0.95)$ cannot be much smaller that
0.06 even for very large $m$. One may conclude that for $z=0.95$ there
is no ``global'' solution in the interval $[-\infty < s < s_0]$.
%%%%%%%%%%%%%%%%%%%%%%%%%%%%%
\begin{figure}
\epsfxsize=8.3cm \epsfbox{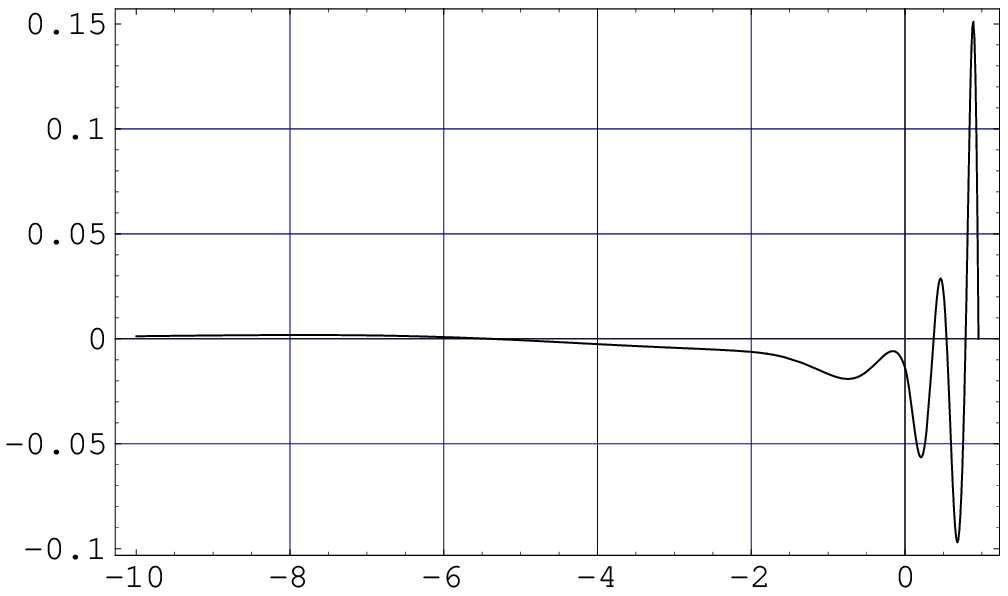}\vskip 0.2cm \epsfxsize=8.3cm
\epsfbox{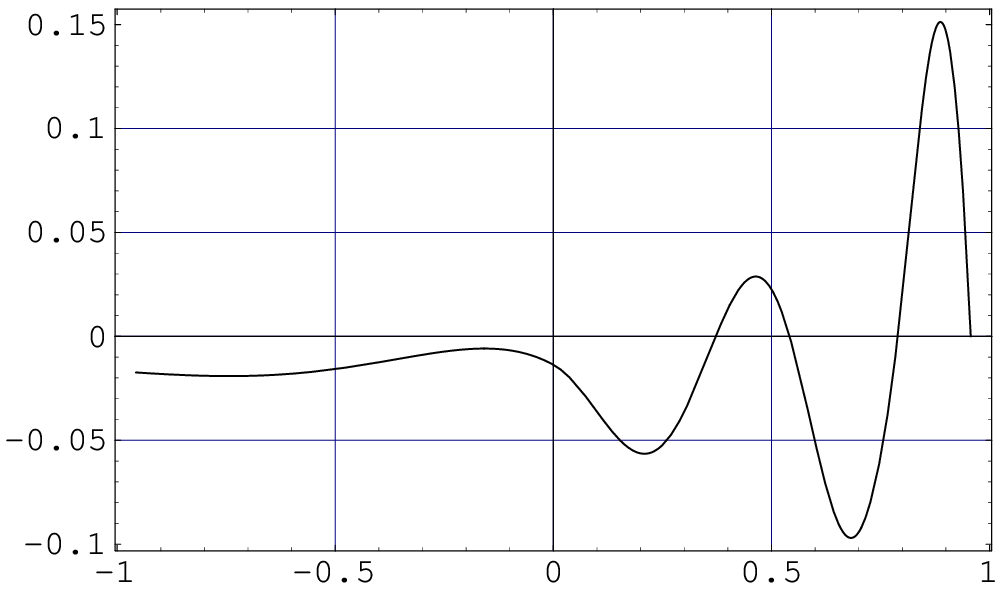}
\caption{Local deviation $\C.D_{{\rm min},5}(s,0.95)$ {\em vs} $s$  
for $\lambda =2$, $b=-0.5$ and $z=0.95$ with optimal set of 10 shape
parameters. Upper panel: region $-10<s<s_0$.  Lower panel: blow up of
the region $-s_0<s<s_0$.}
 \label{f:deviation95}
\end{figure}
%%%%%%%%%%%%%%%%%%%%%%%%%%%%%%%%%%%%
 Nevertheless, the local deviation $\C.D_{{\rm
min},5}(s,0.95)$ shown in Fig.~\ref{f:deviation95} is quite small,
say in the region $s<0$. Moreover, in the region $s<s_*\approx -2$ the
characteristic value of the deviation $\C.D_{{\rm min},5}(s,0.95)$ is
more or less the same as for $z=0.75$ where we have approximate
solution (soliton) in the whole region $s< s_0$. Consequences of
this fact will be discussed later.
%%%%%%%%%%%%%%%%%%%%%%%%%%%%%%%%%%%%%%%%%%%%%%%%%%%%%%%%%%%
\section{ Solitons, quasi-solitons and Asymptotic Multiscaling}
\label{s:solitons}
\subsection{$z$-valleys of the functional $\C.F_{{\rm min,5}}(z)$ }
\label{ss:z-valley}
\begin{figure}
\hskip 0.2 cm \epsfxsize=7.8 cm\epsfbox{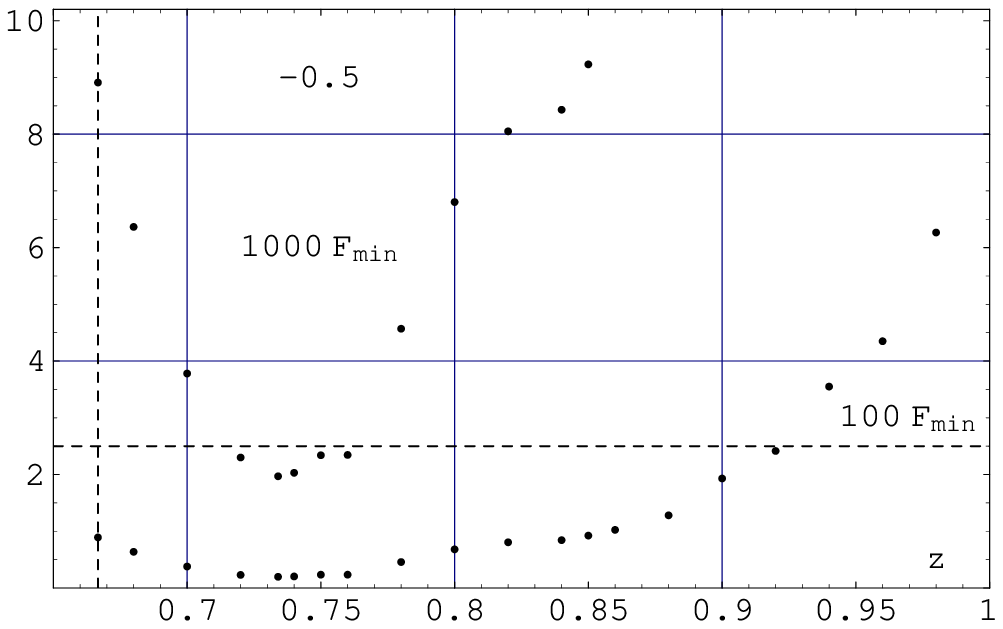}\vskip -0.01cm
\epsfxsize=8cm\epsfbox{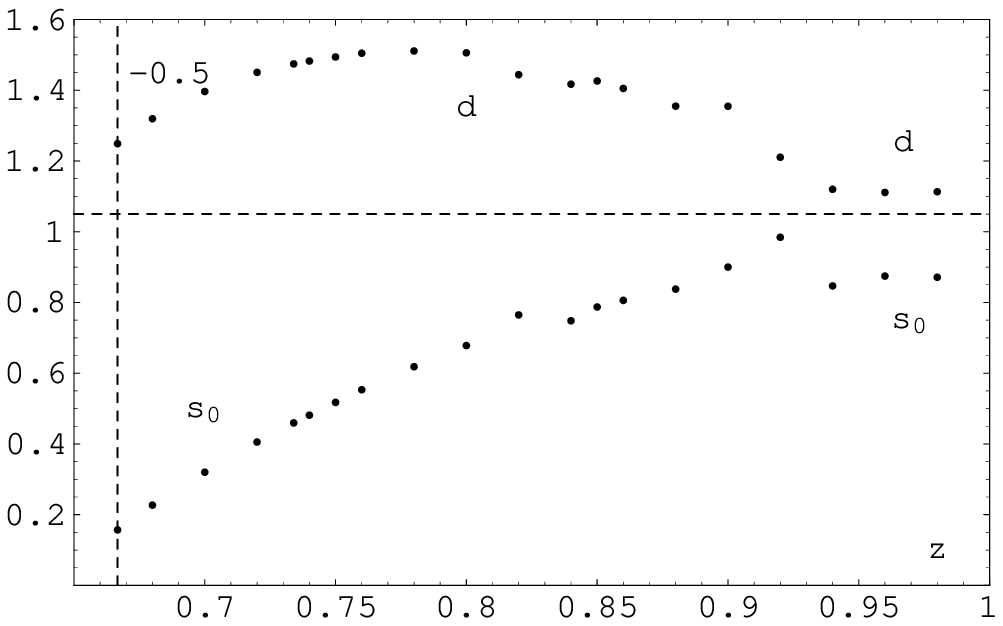}
\caption{The $z$ dependence of $10^3\C.F_{{\rm min,5}}$,   
$10^2\C.F_{{\rm m in,10}}$ (Upper panel) and $s_0$,\, $d$ (Lower
panel) for $\lambda=2$, $b=-0.5$. Vertical dashed lines shows K41
value of $z=2/3$. Horizontal dashed line in upper panel corresponds to
chosen value $10^3 \C.F_*=2.5$.  Points for $10^3\C.F_{{\rm min,5}}$
with $0.72 < z < 0.76$ are below this level.}
\label{f:zs0d-valley50}
\end{figure}

\begin{figure}
\hskip 0.2cm\epsfxsize=8.1cm\epsfbox{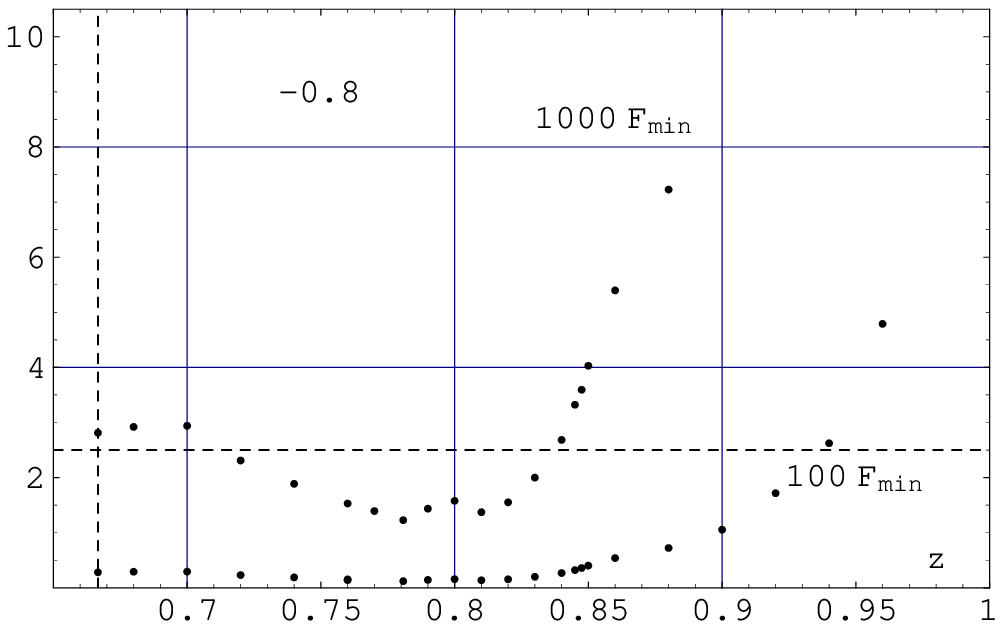} \vskip 0.02cm
\epsfxsize=8.3cm\epsfbox{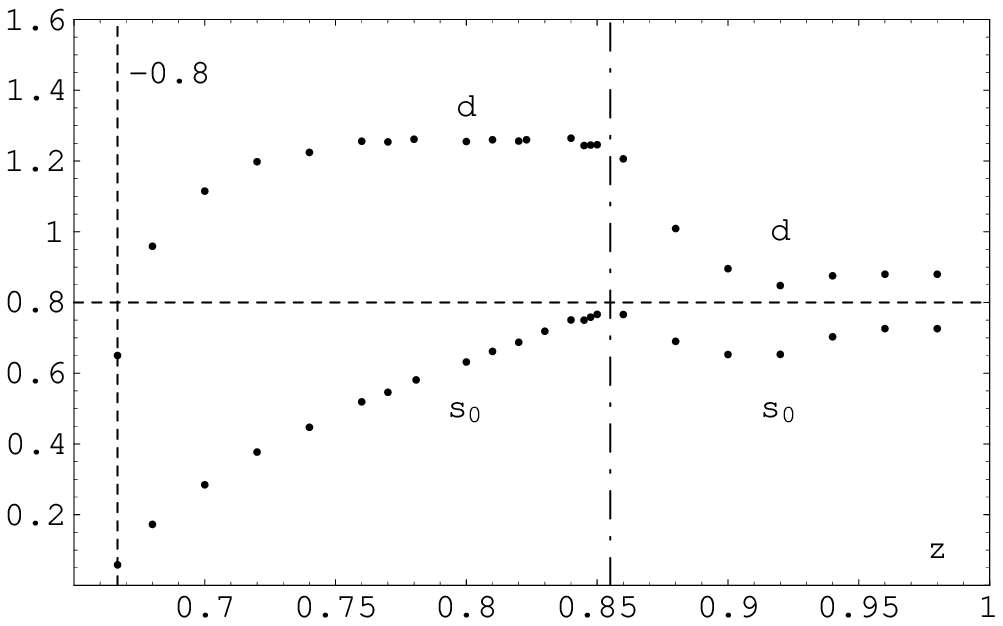}
\caption{The $z$ dependence of $10^3\C.F_{{\rm min,5}}$,   
$10^2\C.F_{{\rm min,5}}$ (Upper panel) and $s_0$,\, $d$ (Lower panel)
for $\lambda=2$, $b=-0.8$.  Vertical dashed line shows K41 value of
$z=2/3$.  Horizontal dashed line in upper panel corresponds to
chosen value $10^3 \C.F_*=2.5$. Points for  $10^3\C.F_{{\rm
    min,5}}$ with $0.71 < z < 0.84$ are below this level. }
\label{f:zs0d-valley80}
\end{figure}

\begin{figure}
\epsfxsize=8 cm ~\hskip 0.3cm \epsfbox{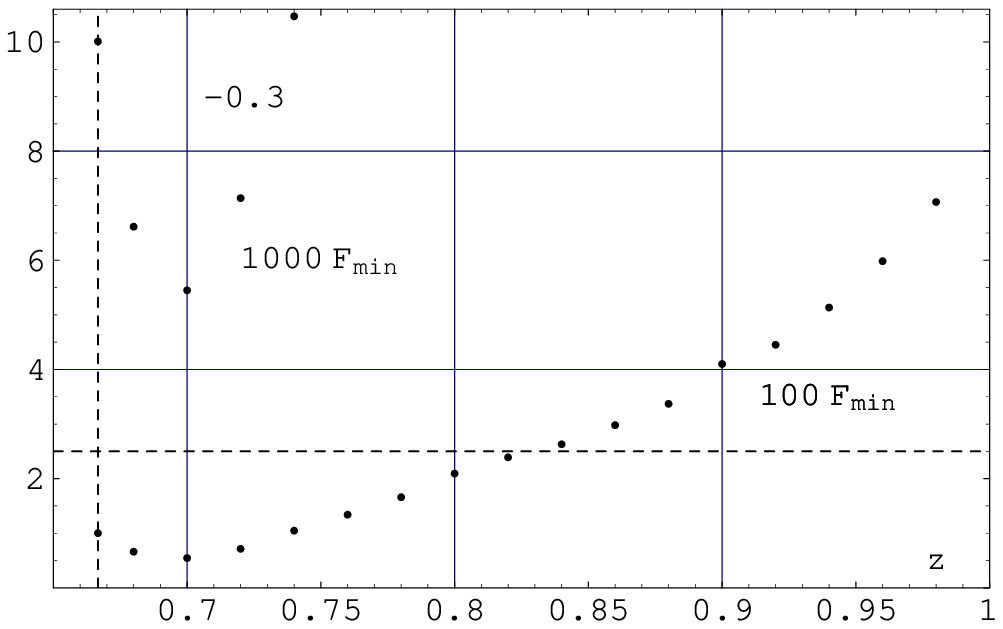}
\vskip 0.02cm
\epsfxsize=8.4cm\epsfbox{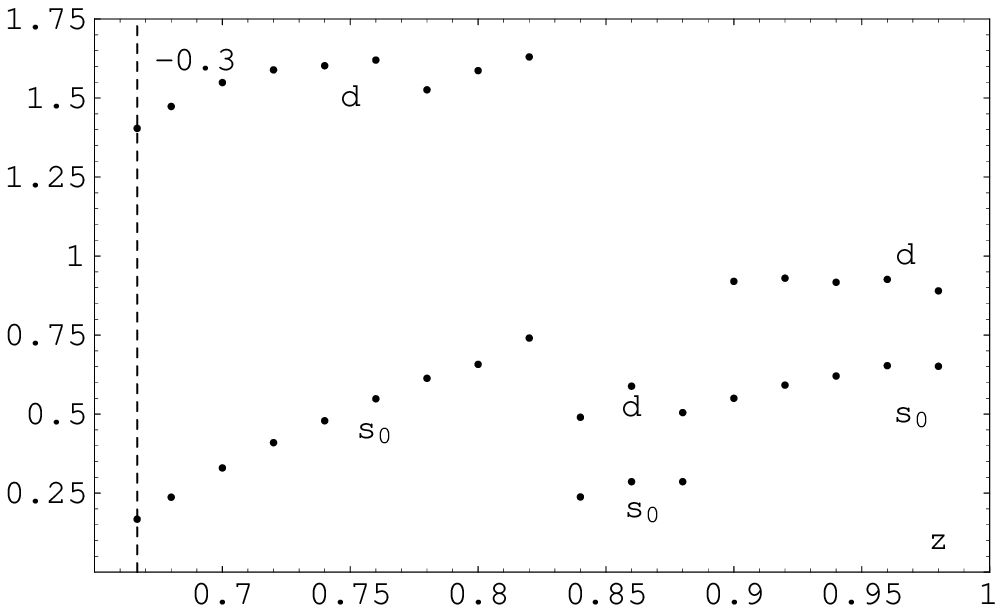}
\caption{The $z$ dependence of $10^3\C.F_{{\rm min,5}}$,   
$10^2\C.F_{{\rm m in,10}}$ (Upper panel) and $s_0$,\, $d$ (Lower
panel) for $\lambda=2$, $b=-0.3$.  The vertical dashed line shows K41
value of $z=2/3$.  Horizontal dashed line in upper panel corresponds
to chosen value $10^3 \C.F_*=2.5$. All points for  $10^3\C.F_{{\rm
    min,5}}$ exceed this level.}
\label{f:zs0d-valley30}
\end{figure}

%%%%%%%%%%%%%%%%%%%%%%%%%%%%%%%%%%%%
In this section we discuss the  $z$-dependence of the functional
$\C.F_{{\rm min},5}(z)$ and of the optimal values of $s_0$ and $d$.
These functions for $b=-0.5$,\ $-0.8$, $-0.3$\ are displayed in
Figs.~\ref{f:zs0d-valley50},\,\ref{f:zs0d-valley80},\,
\ref{f:zs0d-valley30} respectively.

For $b=-0.5$ (Fig.~\ref{f:zs0d-valley50}) the $z$-dependence of the
functional has a minimum $\C.F_{{\rm min,5}}(z)\approx 2\cdot 10^{-3}$
around $z\approx 0.734$. This minimum is quite flat. For example, $
\C.F_{{\rm min,5}}(z)< \C.F_* \equiv 2.5 \cdot 10^{-3}$ for $z$ within
the interval $[z_{\rm min}=0.71, z_{\rm max}=0.76]$.  Note that for
$b=-0.8$, (Fig.~\ref{f:zs0d-valley80}) the same (arbitrary) level
$\C.F_*$ exceeds $\C.F_{{\rm min,5}}(z)$ for $z$ in the wider interval
$[0.71,0.84]$. In the same time for $b=-0.3$
(Fig.~\ref{f:zs0d-valley30}) there are no values of $z$ for which
$\C.F_{{\rm min,5}}(z)< \C.F_*$. A natural interpretation of these
facts

%%%%%%%%%%%%%%%%%%%%%%%%%%%%%%%%%%%%%%%%%%%%%% %%%%%%%%%
%%%%%%%%%%%%%%%%%%%%%%%%%%%%%%%%%%%%%%%%%%%%%%%%%%%%%%%%%

%%%%%%%%%%%%%%%%%%%%%%%%%%%%%%%%%%%%
 
%%%%%%%%%%%%%%%%%%%%%%%%%%%%%%%%%%%
\noindent
 is that at $b=-0.8$ one can meet in the velocity realization (of the
 Sabra model) $u_n(t)$ intense solitons with values of $z$ in a quite
 wide interval (for example in the interval [0.71, 0.84] mentioned
 above), at $b=-0.5$ ``allowed'' interval of $z$ is more narrow (say,
 [0.71, 0.76]) and for $b=-0.3$ one hardly can meet intense solitons
 at all. As we will discuss below this statement is in a quantitative
 agreement with the preliminary results of the numerics~\cite{01Pom}.

Clearly, we are not talking about particular values of $z_{\rm min}$
and $z_{\rm max}$, for example because the boundary level $\C.F_*=2.5
\cdot 10^{-3}$ was chosen arbitrarily. Moreover, the objects $z_{\rm
min}$, $z_{\rm max}$ do not have explicit sense. The actual conjecture
is that the functional $\C.F_{{\rm min},m}(\lambda,b,z)$ is correlated
with a {\em probability} to meet a soliton or quasi-soliton with
given dynamical exponent $z$ in the realization (for given $\lambda$
and $b$): the smaller value of $\C.F_{{\rm min},m}(z)$ (at large
enough $m$), the larger this probability.  From this point of view the
two following scenarios are statistically almost equivalent. The first
one may be called a {\em multi-soliton scenario}: there is a discrete
spectrum of solitons with some close set of exponents $z_{0,1}$,
$z_{0,2}$ in the interval $[z_{\rm min}$, $z_{\rm max}]$.  One can
imagine that this is the case by looking at Fig.~\ref{f:zs0d-valley80}
(upper panel) where the function $\C.F_{{\rm min,5}}(2,-0.8,z)$ has two
minima at $z_{0,1}\approx 0.77$ and $z_{0,2}\approx 0.81$. The second
one will be referred to as a {\em quasi-soliton scenerio} with continuous
$z$-spectrum of quasi-solitons in some (wide) interval of $z$.
%%%%%%%%%%%%%%%%%%%%%%%%%%%%%%%%%%%%%%%%%%%%%%%
\subsection{Local deviations and quasi-solitons}
\label{ss:quasi}
\begin{figure}
 \epsfxsize=8.3cm
\epsfbox{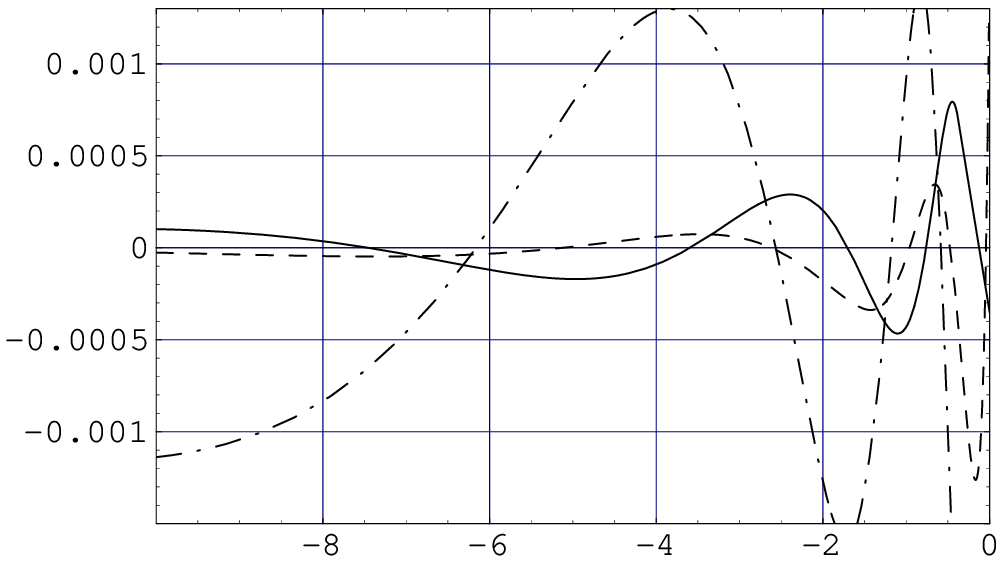}\vskip 0.1cm
\epsfxsize=8.1cm ~\hskip 0.1cm  
\epsfbox{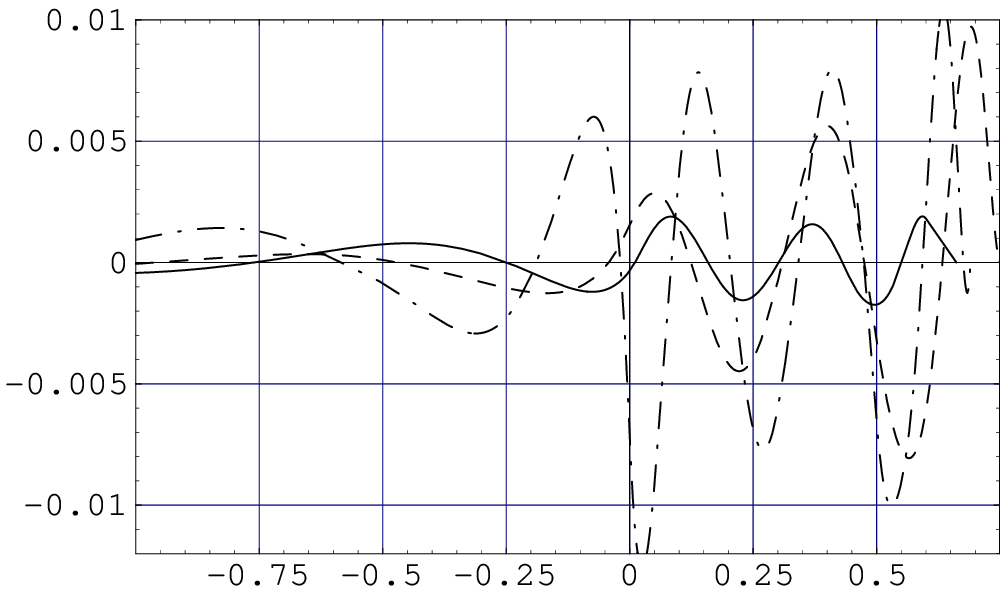}
\caption{Local deviation $\C.D$, Eq.~(\ref{eq:eq-s}) {\em vs} $s$  
for $\lambda =2$ and $b=-0.8$.  Upper panel: region $-10<s<0$.  Lower
panel: region $-1<s<s_0$.  Solid lines: $z=0.81$, dashed lines:
$z=0.84$, dash-dotted lines: $z=0.88$. }
\label{f:deviation80}
\end{figure}

The analysis of the local deviations $\C.D_{{\rm min},m}(s,z)$
presented in this section supports a quasi-soliton scenario of
asymptotic multiscaling. Let us chose $b=-0.8$ for which we found in
Fig.~\ref{f:zs0d-valley80}, (upper panel) the wide deep $z$-valley of
the functional $\C.F_{{\rm min},5}(z)$.  Compare the $s$-dependence of
$\C.D_{{\rm min},m}(s,z)$ for the value of $z=0.81$, which corresponds
to the right local minimum of the functional and for two slightly
larger values of $z=0.84$ and 0.88.  The local deviations $\C.D_{{\rm
min},m}(s,z)$ are plotted in Fig.~\ref{f:deviation80} for $z=0.81$ by
solid lines, for $z=0.84$ by dashed lines and for $z=0.88$ by
dash-dotted lines. The upper panel shows $s$-interval of a left tail
the soliton up to its maximum: $[-10 <s< 0]$. The lower panel shows
$s$-interval around soliton maximum: $[-1<s<s_0]$.

%%%%%%%%%%%%%%%%%%%%%%%%%%%%%
%%%%%%%%%%%%%%%%%%%%%%%%%%%%%
\begin{figure}
 \epsfxsize=8.05cm \epsfbox{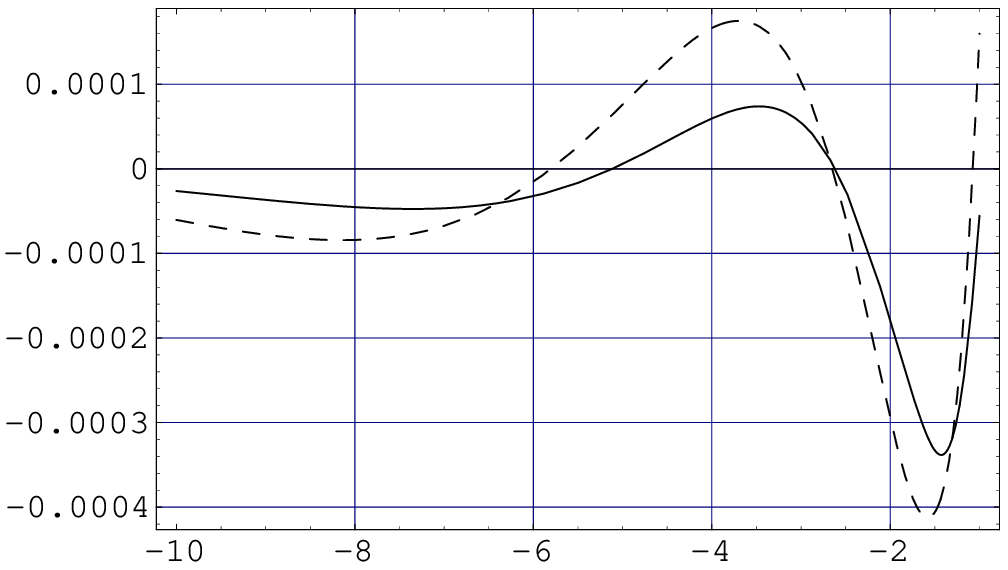}\vskip 0.1cm
 \epsfxsize=8.2cm ~\hskip 0.1cm \epsfbox{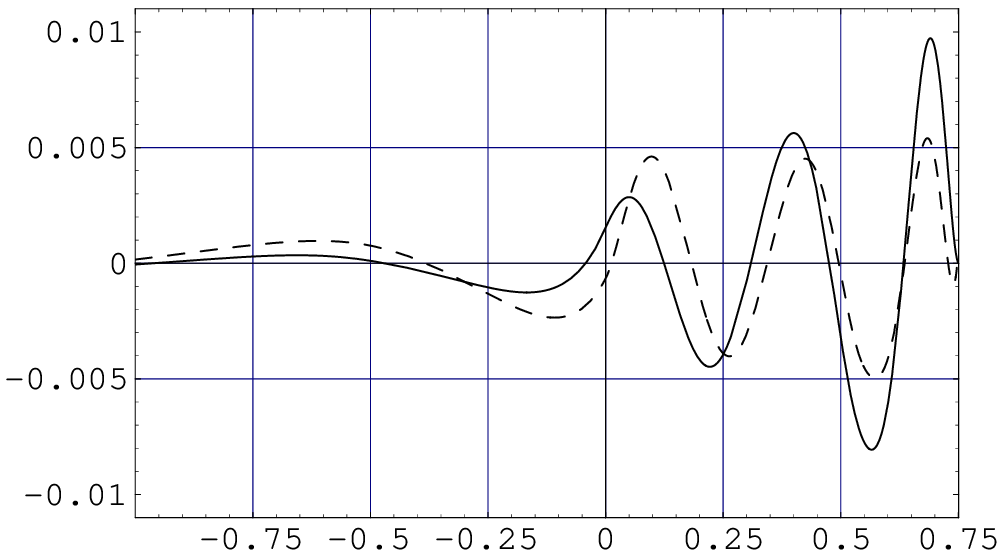}
\caption{Local deviation $\C.D$, Eq.~(\ref{eq:eq-s}) {\em vs} $s$  
for $\lambda =2$, $b$=-0.8 and $z=0.84$. Solid lines-results of
minimization of the functional (\ref{eq:fun}) with unite weight,
dashed lines- the same with the weighed functional (\ref{eq:tfun}),
(\ref{eq:weight}).  Upper panel: region $-10<s<-1$.  Lower panel:
region $-1<s<s_0$.  }
\label{f:dev-comp84}
\end{figure}
%%%%%%%%%%%%%%%%%%%%%%%%%%%%%%%%%%%%
 It was mentioned that the value $z_{0,2}=0.81$ corresponds to the
 right local minimum of the functional.  The local deviation
 $|\C.D_{{\rm min,5}}(s,0.81)|$ is about $2\cdot 10^{-3}$ in the
 $s$-region $[-1, s_0]$, see solid line in Fig.~\ref{f:deviation80},
 lower panel). For $s<-2$ (the upper panel) the local deviation is
 almost 10 times smaller, about $2.5\cdot 10^{-4}$.  The value
 $z=0.84$ does not correspond to $z$-minima of the functional:
 $\C.F_{{\rm min,5}}(0.84)\approx 2\, \C.F_{{\rm min,5}}(0.81)$. The
 larger value of $\C.F_{{\rm min,5}}(0.84)$ is mostly determined by
 the region of positive $s$ where $|\C.D_{{\rm min,5}}(s,0.84)|$
 exceeds 0.01, which essentially larger than the local deviations in
 this region at $z=0.81$, [compare dashed ($z=0.84$) and solid
 ($z=0.81$) lines in the lower panel of
 Fig.~\ref{f:deviation80}]. Unexpectedly, in the region $s<-1$ (the
 upper panel) the local deviation at $z=0.84$ even smaller than that
 at $z= z_{0,2}= 0.81$. In other words, in this $s$-region the reached
 approximation is better for value $z=0.84$ than that for
 $z=0.81$. Dash-dotted lines in Fig.~\ref{f:deviation80} show that the
 local deviation for a $z=0.88$ essentially larger than for $z=0.84$
 and $0.81$.

Let us show that these qualitative results are independent of the
particular form of the functional~(\ref{eq:fun}) which we chose. For
that consider the weighed functional~(\ref{eq:tfun}) with the weight
fuction:
\begin{equation}
  \label{eq:weight}
  W(s)=[(s-s_0)^2+1]^y\,,
\end{equation}
which is of the order of unity in the vicinity of the soliton maximum
and increases toward the front tail as $|s|^{2y}$. At $y=0$ one
recovers the uniform weight $W(s)=1$. For positive $y$ the weight
function emphasize the left tail, {\em i.e.} the region $s\ll 1$. The
value of $y$ has to be smaller than $\case{1}{2}$, otherwise the
$s$-integral in~(\ref{eq:tfun}) diverges. It is reasonable to choose
an intermediate value of $y=1/4$ and to repeate the minimization
procedure.  The resulting changes in the values of $s_0$, $d$ and
shape parameters were minor. For example (for $\lambda=2$, $b=-0.8$,
$z=0.84$) $s_0=0.7506\to 0.7478$, $d=1.2645 \to
1.2618$. Fig.~\ref{f:dev-comp84} compares the local deviation $\C.D$
for unweighed (solid lines) and weighed (dashed lines)
functionals. Again, the difference between these two approaches is
minor.

A natural interpretation of all these facts as is follows: for some
discrete values of $z=z_{0,j}$ (say for $z_{0,1}\approx 0.734$ at
$b=-0.5$ and $z_{0,1}\approx 0.78$, $z_{0,2}\approx 0.81$ at $b=-0.8$)
the variation procedure allows to find better and better approximation
to the solution in the global sense: $\C.F_{{\rm min},m}(z_{0,j})\to
0$ in the limit $m\to \infty$. Clearly in these cases solution $g(s)$
exists also locally: the maximum of the local deviation $\C.D_{{\rm
min},m}(s,z_{0,j})$ also goes to zero in this limit. In means that
there are self-similar solitons in the shell model with scaling
exponents $z_{0,j}$.  For other values of $z$ there is no global
solution. It means that for $m\to \infty$ limit of $\C.F_{{\rm
min},m}$ is finite, but may be small enough. Nevertheless for a wide
region of $s<s_*$ (in the example $z=0.75$, $b=-0.5$ the value of $s_*
\approx -2$) the local deviation $\C.D_{{\rm min},m}(s,z)$ is very
small and equation~(\ref{eq:eq-s}) may be ``almost solved''. We may
say that the variation procedure finds ``quasi-solutions'', which
approximate the equations of motion with very good accuracy starting
from the ``time of appearance of a soliton ($s\to -\infty$) until some
time $s_*$. The value of $s_*$ may be close to the time of the soliton
maximum. Remember, that in the real shell systemintense events form
in a random background of small fluctuations where one may meet
configurations corresponding to initial conditions of these
``quasi-solutions''.  In that case the system begins to evolve along the
quasi-solutions up to some time about $s<s_*$ where the quasi-solution
stops to approximate well the self-similar equation of motion.  These
peaces of trajectory are referred to as {\em quasi-solitons}.  After
the time $s_*$ trajectory has to deviate from the quasi-solitons,
presumably exponentially fast.  Quasi-solitons look like the  front part
of solitons and may include their maxima.

 The main message is that {\em the quasi-solitons have a continuous
 spectrum of $z$ which vary a lot around $z_0$. Therefore their
 contribution to the asymptotic multiscaing may be even more important
 than the contributions of trajectories in the vicinity of the soliton
 with fixed scaling exponent $z_0$. Moreover in some region of
 parameters} [in Sabra model at $\lambda=2$ for $b>b_{\rm cr}\approx
 -( 0.5- 0.4)$] {\em there are no solitons and the quasi-solitons
 provide dominant contribution to the asymptotic multiscaling}.

This picture is consistent with the preliminary direct numerical
simulation of the Sabra shell model\cite{01Pom,01LPP} where it was
observed: i) at $b=-0.8$ self-similarintense events with different
rescaling exponents (see wide deep minimum of $\C.F_{{\rm min},m}(z)$
in Fig.~\ref{f:zs0d-valley80}); ii) at $b=-0.5$ self-similar events
with a very narrow region of $z$ (see not so deep minimum of
$\C.F_{{\rm min},m}(z)$ in Fig.~\ref{f:zs0d-valley50}); and iii) no
solitons at $b=-0.3$ (there is no deep minimum in
Fig.~\ref{f:zs0d-valley30}).

\section{Conclusion}
\label{s:concl}
$\bullet$ I have suggested a variation procedure (basic functional and
trial functions) for an approximate solution of equations for the 
self-similar solitary peaks (solitons) in shell models of
turbulence. The rational functions with 10 shape parameters
approximate solitons in the Sabra shell model with relative accuracy
 $0(10^{-3})$.

$\bullet$ For the standard set of parameters ($\lambda=2$, $b=-0.5$)
the dynamical exponent $z_0\approx 0.734$ found in the paper agrees
within the error bars with the experimental value of $z_0\approx
0.75\pm 0.02$~\cite{01LPP}.

$\bullet$ The variation procedure allows to find trajectories of the
system which are very close to the self-similar solitons during an
interval of time from $-\infty$ up to some time in the vicinity of the
soliton maximum or even after it. These trajectories, called
quasi-solitons have continuous spectrum of the dynamical scaling
exponents and may provide dominant contribution to asymptotic
multiscaling.  The discovered features of quasi-solitons for
$b=-0.8\,,\ -0.5$, and $-0.3$ at $\lambda=2$ allows me to rationalize
the preliminary numerical observations\cite{01LPP,01Pom} of asymptotic
multiscaling for various values of $b$ in the Sabra model.

$\bullet$ This paper may be considered as a first step toward a
realistic statistical theory of asymptotic multiscaling in shell
models of turbulence which will account for a wide variety of
trajectories of the system in the vicinity of quasi-solitons.  In
particular, one can apply the variation procedure to find possible
complex solitons and quasi-solitons with non-trivially rotating
phases. Such objects may be important in quantitative description of
high-level intermittency when the parameter $b$ approaches the
critical value $-1$. Analytical form of self-similar trajectories of
the system found in this paper may help one to analyze stability of
nearby trajectories of the system. This is useful for describing how
the system approaches and later escapes a vicinity of quasi-solitons
in order to describe their role in statistics ofintense but not
solitary events.

\vskip 0.5cm

I hope that the analysis of dynamics ofintense events in the Sabra
model, presented in the paper will help in futher understanding of
asymptotic multiscaling along similar lines. These will be also useful
for further progress in the problem of anomalous scaling in
Navier-Stokes turbulence at least on a qualitative and may be on
semi-quantitative level.

 %%%%%%%%%%%%%%%%%%%%%%%%%%%%%%%%%%%%

\acknowledgments It is a pleasure to acknowledge numerous elucidative
conversations with I. Procaccia and A. Pomyalov which contributed to
this paper.  This work has been supported by the Israel Science
Foundation.

%%%%%%%%%%%%%%%%%%%%%%

\end{multicols}
\end{document}